\documentclass[lettersize,journal]{IEEEtran}

\usepackage{amsmath,amsfonts,amssymb,textcomp}

\usepackage{graphicx}
\usepackage[caption=false,font=normalsize]{subfig}

\usepackage{array}
\usepackage{booktabs}
\usepackage{tabularx}

\usepackage{algorithm}
\usepackage{algpseudocode}

\usepackage{url}
\usepackage{verbatim}
\usepackage{xcolor}
\usepackage[skip=0.5ex]{caption}
\usepackage{stfloats}
\usepackage{cite}
\usepackage[utf8]{inputenc} % specify the input encoding
\usepackage[T1]{fontenc}    % specify the font encoding
\usepackage{amsmath}
\usepackage{amssymb}
\usepackage{multirow}
\usepackage{graphicx}
\usepackage{subfig} 
\begin{document}

\title{Signal Watermark on Large Language Models}

        % <-this % stops a space


\author{
    Zhenyu Xu, Victor S. Sheng\\
    Department of Computer Science, Texas Tech University\\
    \{zhenxu, victor.sheng\}@ttu.edu
}
\maketitle
\begin{abstract}
As Large Language Models (LLMs) become increasingly sophisticated, they raise significant security concerns, including the creation of fake news and academic misuse. Most detectors for identifying model-generated text are limited by their reliance on variance in perplexity and burstiness, and they require substantial computational resources. In this paper, we proposed a watermarking method embedding a specific watermark into the text during its generation by LLMs, based on a pre-defined signal pattern. This technique not only ensures the watermark's invisibility to humans but also maintains the quality and grammatical integrity of model-generated text. We utilize LLMs and Fast Fourier Transform (FFT) for token probability computation and detection of the signal watermark. The unique application of signal processing principles within the realm of text generation by LLMs allows for subtle yet effective embedding of watermarks, which do not compromise the quality or coherence of the generated text. Our method has been empirically validated across multiple LLMs, consistently maintaining high detection accuracy, even with variations in temperature settings during text generation. In the experiment of distinguishing between human-written and watermarked text, our method achieved an AUROC score of 0.97, significantly outperforming existing methods like GPTZero, which scored 0.64. The watermark's resilience to various attacking scenarios further confirms its robustness, addressing significant challenges in model-generated text authentication.
\end{abstract}


\section{Introduction}
With the advancement of Large Language Models (LLMs) such as ChatGPT \cite{chatgpt}, their capabilities have become increasingly sophisticated, enabling them to perform a wide range of tasks with remarkable efficiency and accuracy. Demonstrating formidable abilities, these models can answer scientific and mathematical queries, understand historical events and biographies, and offer assistance with daily life. Even though language models may occasionally produce illusions or incorrect answers, the potential and impact of LLMs remain significant. However, this enhanced functionality also presents new challenges in terms of security and potential misuse. For instance, there's a risk of these models being used to create fake news or media rumors, and students might leverage them to generate assignments, thereby impacting academic fairness. Consequently, there is a growing need for reliable methods to identify model-generated text to effectively address and mitigate these emerging concerns.
\begin{figure}[!h]
    \centering
    \includegraphics[width=\linewidth]{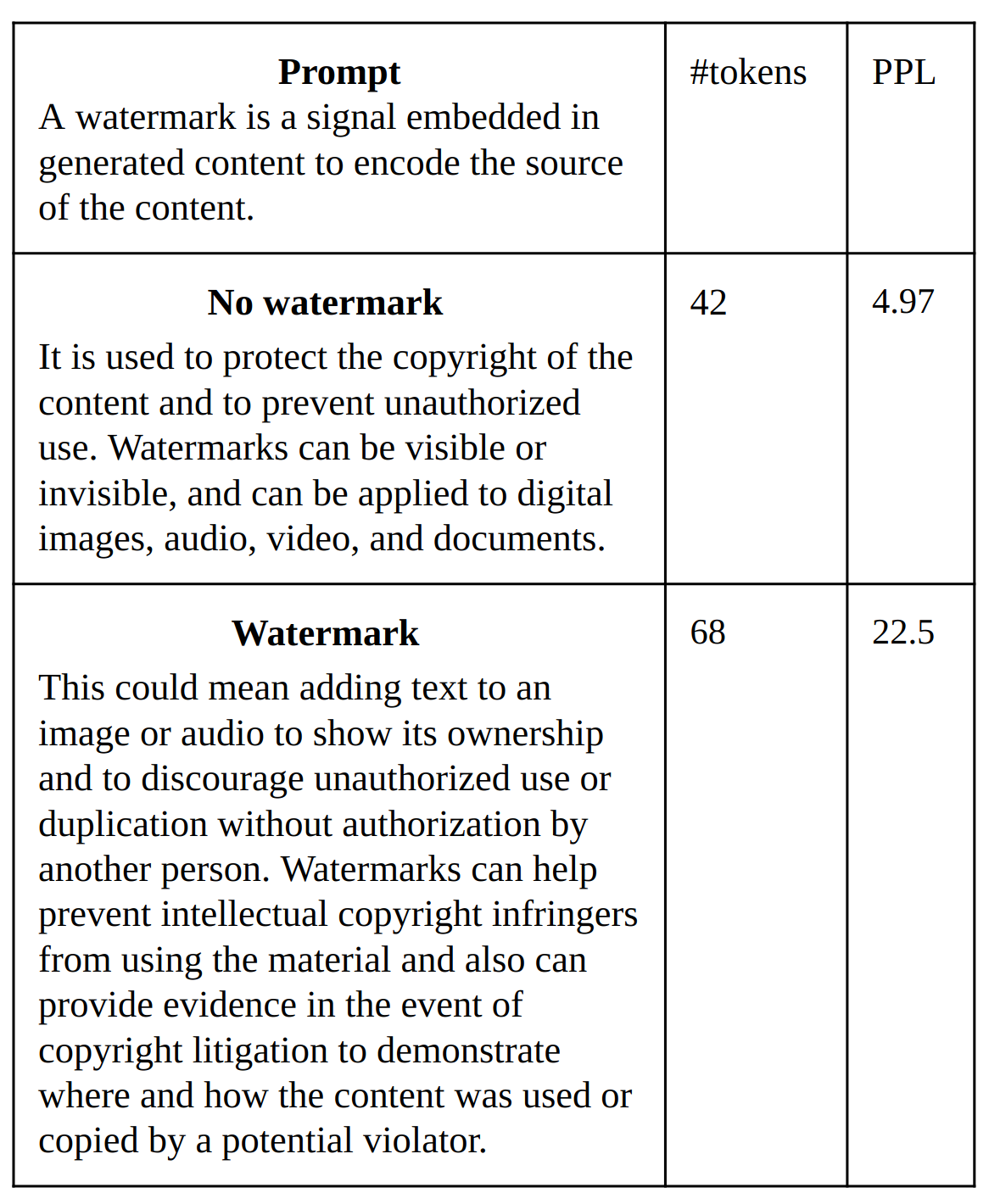}
    \caption{This figure illustrates two samples: one without a watermark and the other embedded with a watermark, both generated by the text-davinci-003 \cite{td003} model. Both samples are produced by the text-davinci-003 model. The Perplexity (PPL) scores, calculated by the same model. It indicate that even with the watermark embedded, the text maintains a regular text quality, similar to high-quality human-written texts.}
    \label{fig:W&NW}
\end{figure}

Numerous open-source and commercial tools have been developed to detect model-generated text, such as GPT2-Detector \cite{Solaiman2019}, RoBERTa-QA \cite{Guo2023}, DetectGPT \cite{Mitchell2023}, GPTZero \cite{GPTZero}, Writer \cite{Writer}, and others. However, there are three significant limitations to these methods. First, their probability predictions cannot provide conclusive certainty and should only be regarded as a reference. Second, most current approaches rely on methods that differentiate human text from model-generated text based on variations in perplexity and burstiness. Third, methods like DetectGPT and RoBERTa-QA, which require multiple text-generation process and model retraining, consume substantial computational resources and time when detecting model-generated text. As LLMs advance, the generated text increasingly resembles human writing, making these methods based on variance in perplexity between human and model-generated text less applicable. Therefore, embedding watermarks in text generated by LLMs presents a faster, more stable, and reliable method of verification. 

Watermarks in Large Language Models need to meet three key criteria: First, the watermark should be imperceptible to humans. Second, it must maintain the text's quality, ensuring semantic accuracy, context relevance, and grammatical integrity. Third, the watermark should exhibit robustness against various attacks. To address these requirements, we propose a novel watermarking technique for texts generated by LLMs, embedding watermarks through signal waveform patterns. This method employs a token sampling process, governed by specific rules, to insert watermark tokens subtly into the text, thereby preserving its coherence. This approach successfully satisfies all three criteria. Figure \ref{fig:W&NW} provides a side-by-side comparison of model-generated texts, one with and one without the watermark, showcasing the technique's effectiveness in maintaining text quality while embedding the watermark.

\begin{figure*}[!ht]
    \centering
    \includegraphics[width=\linewidth]{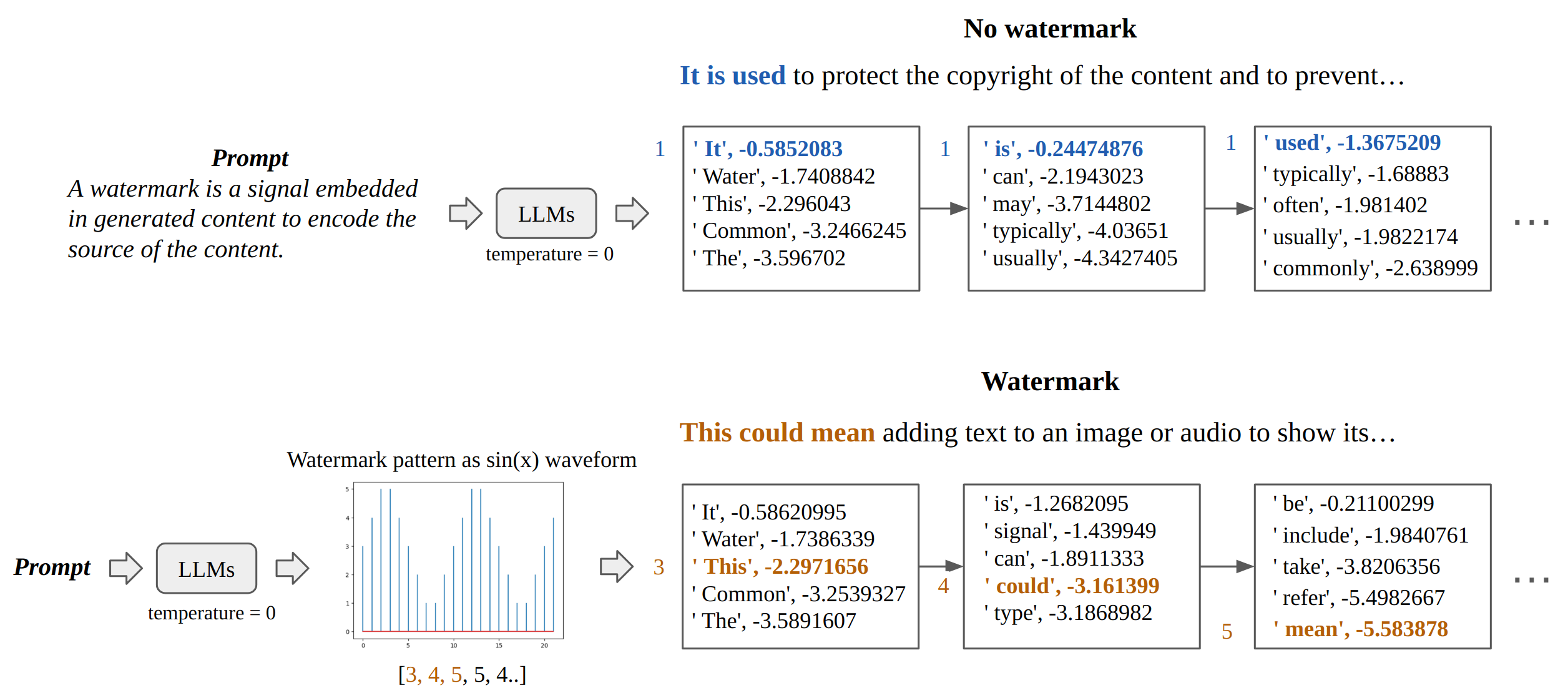}
    \caption{This figure illustrates an example of embedding a signal watermark while generating text by LLMs. The black box displays our top 5 token candidates pool, and  the temperature is set to 0. When the model generates text with no watermark, it consistently selects the token with the highest probability as the next token when no specific pattern is followed, as indicated in blue. Meanwhile, when the model generates watermark text using the same prompt, we have a pre-defined pattern sampling from sin(x), dictating the rank of the token to be selected from token candidates pool. Each time a next token is generated, it was chosen according to this pattern, shown in orange. Higher temperatures would perturb this sampling process, leading the model to select tokens that deviate from the most probable one. For examples of text generation at different temperatures, please see Appendix A.}
    \label{fig:WProcess}
\end{figure*}

Before delving into the concept of signal watermarking, it is crucial to grasp the fundamental text generation process employed by Large Language Models (LLMs) like those in the GPT (Generative Pre-trained Transformer) series \cite{radford2018improving}\cite{radford2019language}\cite{brown2020language}. These models generate text from a given prompt by sequentially predicting one token at a time. This method is known as auto-regressive text generation. In this approach, each time a token is generated, it is concatenated with the entire sequence of the initial prompt and previously generated tokens to form a new input. This updated input is then used by the model to predict the next token. With each prediction, the generated text extends. For each token prediction, the model produces a pool of potential token candidates, each with an assigned log probability. In this context, log probabilities are typically negative numbers; the closer this value is to zero, the higher the probability of that token being the next choice in the sequence.

Similarly, We use Large Language Models for Token Probability Re-computation, a process where each token in a text is analyzed to generate a pool of candidate tokens and their log probabilities. This is done in an auto-regressive manner: the model sequentially inputs each token of the text and leverages its predictive capabilities to assess the probability distribution for each token's position. In each step, the model considers the cumulative input sequence up to that position and predicts the likelihood of potential next tokens, assigning a probability score to each. These candidates are then ranked by their log probabilities in descending order, with higher values indicating a higher likelihood of being the appropriate successor in the context. Furthermore, we can determine the rank of the actual subsequent token within its candidate pool. It’s crucial to note that this process focuses exclusively on the model-generated text, independent of any initial prompt. Additionally, we do not select any tokens from the candidate pool as input. 
\begin{figure*}[!t]
    \centering
    \subfloat[Token Probability Re-computation]{ % Added description after \subfloat
        \includegraphics[width=2\columnwidth]{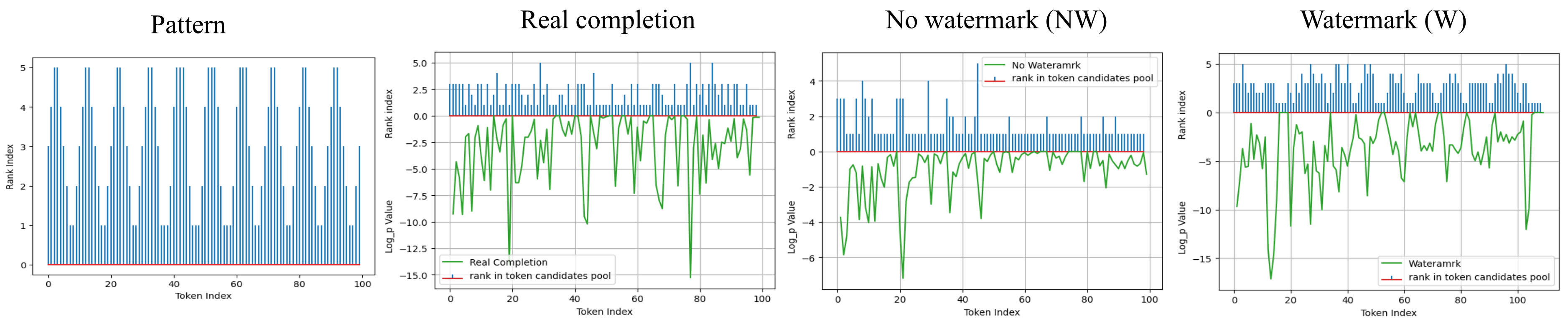}
        \label{fig:RWa}
    }
    
    \subfloat[Watermark Detection]{ % Added description after \subfloat
        \includegraphics[width=2\columnwidth]{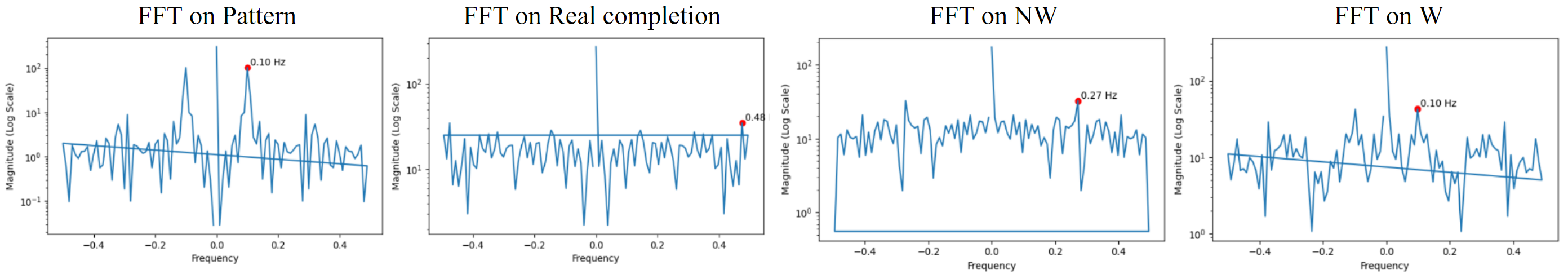}
        \label{fig:RWb}
    }
    \caption{
We perform Token Probability Re-computation on model-generated text and utilize Fast Fourier Transform (FFT) for watermark detection. 'Real completion' refers to human-written text following the prompt, while 'No watermark' and 'Watermark' are texts generated by the model with same prompt. In (a) part of this figure, Pattern represents in a sinusoidal signal with a period of 10 and 10 samples per period. The upper blue waveform of Real completion, No Watermark (NW), and Watermark (W) shows the rank of the actual token within its candidate pool after the re-computation process. The lower green waveform indicates the log probability value of the current token. In the 'Watermark' plot, we can observe a periodic waveform similar to the pre-defined pattern. In (b) part, it presents the frequency spectrum obtained from applying FFT to the blue waveform in (a). In the FFT on W spectrum, a peak frequency of  signal watermark is observed, which corresponds to the pre-defined pattern frequency in the FFT on pattern, highlighted by a red dot at the position of 0.1. In contrast, the spectrum from FFT on Real completion and FFT on NW do not show the same peak frequency. This method of calculating the frequency spectrum via FFT enables the detection of the watermark's presence.}
\label{fig:RW}
\end{figure*}
Our research is grounded in a fundamental hypothesis: when a language model auto-regressively generates text from a given prompt, the ranking of each token in the candidate pool during generation closely matches its ranking in a subsequent Token Probability Re-computation. Note that these candidate pools are sorted in descending order by their probabilities. In other words, for semantically similar input sequences, the model's predicted probability distribution for the next token should be nearly identical, resulting in significant similarities between the candidates pool during the generation process and that obtained in subsequent re-computations. For example, when texts are generated using the top-1 candidate tokens, meaning the most probable next token is consistently chosen. In such cases, we will observe that these tokens generally maintain their top-ranking positions in the candidate pools during later re-computations. We empirically validate this hypothesis and find that it holds true across various Large Language Models.

We have developed a Signal Watermark method that capitalizes on the consistency of token rankings observed during text generation and subsequent re-computation by language models. In this approach, a signal watermark, based on a pre-defined signal pattern, is embedded into the text during its generation. This pattern influences the choice of ranked token candidates at each step of generating the next token. Figure \ref{fig:WProcess} visually demonstrates this process, contrasting texts generated with and without the watermark. During the re-computation phase, we examine the positions of actual tokens within the token candidate pools. This examination is aimed at detecting a pattern that matches the pre-defined pattern. Furthermore, we use Fast Fourier Transform (FFT) \cite{FFT} to calculate the frequency of the embedded signal watermark, enabling the precise identification of the watermark. A key aspect of the FFT method is that identical patterns always yield the same frequency, providing a reliable method to confirm the presence of our specific watermark. Figure \ref{fig:RW} presents an example of Token Probability Re-computation and Watermark Detection Process on texts from three different sources.

Our main contributions are:
\begin{itemize}
\item We have identified and empirically validated the hypothesis that Large Language Models maintain consistent token ranking patterns in both text generation and subsequent re-computation phases.
\item We developed an innovative technique for embedding a signal watermark into text generated by Large Language Models.
\item We introduced the application of Digital Signal Processing (DSP) techniques to improve the detection of watermarks in text generated by these models.
\item Our approach obviates the need for additional model training, streamlining the watermarking process and reducing computational overhead.
\item In the watermark attacking section, we have demonstrated that our signal watermark is resilient to remove.
\end{itemize}

\section{Related Work}
\subsection{Pre-trained Large Language Models}
In the domain of natural language processing (NLP), the advent of pre-trained large language models represents a pivotal milestone. Google's BERT (Bidirectional Encoder Representations from Transformers), pioneered by Devlin et al. \cite{Devlin2019BERT} in 2019, heralded a new era in NLP with its transformer-based, bidirectional approach, significantly enhancing the understanding of language context. Building upon this, RoBERTa, a brainchild of Facebook AI, developed by Liu et al. \cite{Liu2019RoBERTa}, refined BERT's framework, yielding superior performance across various NLP tasks. OpenAI's seminal contributions, GPT-2 and GPT-3, authored by Radford et al. \cite{Radford2019GPT2} and Brown et al. \cite{Brown2020GPT3} respectively, forged new paths in generative language modeling. These models exhibited unparalleled proficiency in generating coherent and contextually nuanced text, establishing unprecedented standards for generative models. Meta, formerly known as Facebook, also made notable advancements with OPT (Open Pre-trained Transformer) and LLaMA (Large Language Model Meta AI). OPT, introduced by Zhang et al. \cite{Zhang2022OPT} in 2022, underscores a commitment to ethical AI development, addressing critical issues like bias and transparency. Concurrently, LLaMA, revealed by Smith et al. \cite{Smith2022LLaMA} in 2022, concentrates on model efficiency, stretching the limits of computational resource utilization.

\subsection{Detection for Model-generated Text}
\subsubsection{Watermark Methods}
Recent advancements in watermarking techniques for language models have been crucial in ensuring text authenticity and model protection. Abdelnabi and Fritz \cite{Abdelnabi2021} pioneered the Adversarial Watermarking Transformer (AWT), marking a significant step in embedding binary messages into text and demonstrating resilience against multiple attacks. He et al. \cite{He2022} developed CATER, a conditional watermarking framework that secures text generation APIs against imitation attacks while preserving word distribution integrity. Zhao, Li, and Wang \cite{Zhao2022} introduced Distillation-Resistant Watermarking (DRW), a method effective in protecting NLP models from distillation theft and maintaining accuracy. Kirchenbauer et al. \cite{Kirchenbauer2023} proposed a framework for embedding watermarks in language models that are detectable with minimal impact on text quality. In a similar vein, Zhao, Wang, and Li's GINSEW technique \cite{Zhao2023} innovatively embeds secret signals in the probability vector of decoding steps, enhancing IP infringement detection. Peng et al. \cite{Peng2023} developed EmbMarker for EaaS, embedding watermarks in trigger words for effective copyright verification. Yoo et al. \cite{Yoo2023} explored robust multi-bit watermarking, focusing on error resistance and enhanced payload. Lastly, Li et al. \cite{Li2023} introduced a novel strategy for embedding watermarks in the weight quantization process of large language models, ensuring model security without compromising operational modes. Together, these contributions represent a diverse and evolving approach to watermarking in the rapidly advancing field of language models.

\subsubsection{Detectors} 
The advancement in language models has prompted the development of various detection tools for identifying model-generated text. Gehrmann \cite{Gehrmann2019} developed GLTR, a tool to assist non-experts in detecting texts generated by models. It uses statistical methods to identify generation artifacts and has been shown to improve human detection rates of fake text significantly. GPT2-Detector developed by OpenAI \cite{Solaiman2019} is a RoBERTa base fine-tuned with outputs from the 1.5B-parameter GPT-2 model. Its primary function is to predict if text was generated by a GPT-2 model. Mitchell et al. \cite{Mitchell2023} introduced DetectGPT, a method for zero-shot detection of machine-generated text based on the curvature of the language model's probability function. This approach does not require training a separate classifier or collecting a dataset of real or generated texts. Guo et al. \cite{Guo2023} focused on understanding how ChatGPT compares to human experts. They developed a RoBERTa Detector and conducted extensive experiments to effectively detect whether text is generated by ChatGPT or humans, providing a comprehensive analysis of the differences between machine and human-generated content. Alongside academic research, there are several commercial tools that have emerged to address the challenge of detecting model-generated text. These include GPTZero \cite{GPTZero}, Writer \cite{Writer}, AITextClassifier \cite{AITextClassifier}, Copyleaks \cite{Copyleaks}, and Sapling \cite{Sapling}. These commercial tools complement the academic models by providing accessible, user-friendly platforms for a wide range of users.

\section{Background}
\subsection{Sampling in Large Language Models} Sampling refers to the method by which an LLM selects the next token in a sequence during generation. When generating text, an LLM doesn't merely choose the most probable next token. Instead, it samples from a probability distribution of potential next tokens, known as the token candidates pool. Each token's probability, determined by the model, reflects its contextual fit semantically and syntactically. The sampling process can be influenced by parameters like temperature and top-k:

\textbf{Temperature} affects how likely it is for a Large Language Model to choose different tokens. When the temperature is high, the model is more likely to pick less common tokens, making the text more varied and less predictable. A low temperature means the model prefers more likely tokens, often making the text more predictable and coherent. Hence, since the temperature parameter alters the likelihoods within the token candidates pool, it can potentially affect the accuracy of watermark detection. 

\textbf{Top-k Sampling} doesn't alter the probability distribution itself. Instead, it limits the pool of token candidates from which the model samples. By selecting only the top 'k' tokens with the highest probabilities, top-k sampling constrains the choice to a smaller, more probable set of tokens. However, unlike temperature, top-k sampling does not alter the underlying probabilities of the tokens. Given that our token candidates pool is ranked by log probability, the implementation of top-k sampling does not influence our pool while generating watermark text. 

\subsection{Fast Fourier Transform} Fast Fourier Transform (FFT) is a fundamental algorithm in the field of Digital Signal Processing (DSP), widely recognized for its efficiency in converting time-domain signals into their frequency-domain counterparts. This transformation is crucial in DSP, as it allows for the detailed analysis of signal characteristics that are not immediately apparent in the time domain. By breaking down complex signals into individual frequency components, FFT facilitates a deeper understanding of the signal's structure and behavior. In the context of our Signal Watermark method, FFT plays a vital role. It is used to examine and identify the frequency of signal watermarks embedded into model-generated text.

\section{Approach}

This section outlines our approach to embedding and detecting signal watermarks in text generated by Large Language Models. Our method integrates concepts from sampling of LLMs and digital signal processing to create a robust and stealthy watermarking system.

\subsection{Signal Watermark}

In the proposed approach, watermarking in text generated by Large Language Models is achieved through a methodically structured process. Initially, a pattern is generated from a periodic sinusoidal signal:
\begin{equation}
    \text{signal\_wave} = \sin(nx + \phi)
    \label{eq:signal_wave}
\end{equation}

This signal's amplitude is scaled to a specific range for token selection, pivotal in the subsequent text generation process. Such as 1 to 5 (other ranges like 1 to 3 or 1 to 4 can also be chosen), and then the pattern is sampled. Each sample point in the text corresponds to a token position, with the signal's current amplitude serving as the rank for selecting tokens from the sorted candidates pool. For example, a sampled \( \sin(x) \) signal yield a pattern such as [3, 4, 5, 5, 4, 3, 2, 1, 1, 2, 3, 4, 5, 5, ...], determining the length of the pattern to match the desired token length. Sampling the scaled wave to create a pattern:
\begin{equation}
    \text{pattern}(n) = \text{signal\_wave}(nT_s) \times \frac{\text{range\_max}}{2} + \text{offset}
\end{equation}
Where pattern(n) is the discrete representation of the sampled signal at the $n^{\text{th}}$ sample. \text{signal\_wave}($nT_s$) represents the value of the continuous time signal \text{signal\_wave} at the sampling instant $nT_s$, $n$ is the sample index and $T_s$ is the sampling period. The terms $\frac{\text{range\_max}}{2}$ and offset are used to scale and adjust the sampled values, respectively.

The watermark embedding process begins with the introduction of a prompt to the model, which responds by generating a set of candidates consisting of the top-5 potential next tokens and their corresponding log probabilities, ranked in descending order. From this set, a token is selected based on its alignment with the appropriate rank indicated by the generated pattern, and this token is then integrated into the generated text. This selection is executed iteratively for each token position \( i \) in the generated text. Specifically, the model's output logits are adjusted according to the specified temperature parameter and subsequently normalized through a softmax function to yield a probability distribution, which guides the selection of the next token. In this case, the softmax function converts $\text{logits}_i$ into a probability distribution $\text{probs}_i$ that corresponds to each token in the vocabulary. Each probability value $\text{prob}_j$ occurs in pairs with a particular token $\text{token}_j$, which together form the \text{token candidates pool}. This process is delineated as follows, where $N$ denotes the total number of all possible subsequent tokens given by model:
\begin{equation}
    \text{probs}_i = \text{softmax}(\frac{\text{logits}_i}{\text{temperature}})
\end{equation}

\begin{equation}
    \text{token\_candidates\_pool}_i = \bigcup_{j=1}^{N} \{ (\text{token}_j, \text{prob}_j) \}
\end{equation}

From the \text{token candidates pool}, the top-5 most probable tokens are extracted and ranked in decrease order by log probability, denoted as $candidate\_set$. The token that aligns with the rank indicated by the sinusoidal pattern at the current position $i$ is then selected. This selection process is formulated as:

\begin{equation}
    \text{selected\_token}_i = \text{candidate\_set}_i[\text{pattern}[i] - 1]
\end{equation}

This selected token is then appended to the accumulated generated text. Subsequently, the process iterates by combining the initial prompt with the accumulated generated tokens to predict the next token. This iterative approach embeds a watermark into the LLM-generated text while ensuring the text remains coherent and contextually relevant, thus preserving its natural linguistic flow.

\subsection{Token Probability Re-computation}

Token Probability Re-computation involves recalculating the token candidates and their log probabilities for a given segment of text. This process, fundamental to our research method, adopts an auto-regressive approach for sequential text input into the model. It allows for the prediction and analysis of each token's probability distribution and its relative position within the candidate set. During the Re-computation process, models' temperature is always set to 0. The algorithm describes as follows:

\begin{algorithm}
\caption{Token Probability Re-computation Process}
\label{Token Probability Recomputation Process}
\begin{algorithmic}[1]
\State \textbf{Input:} Text segment $T$, Model $M$
\State \textbf{Output:} Rank indices for each token in $T$
\State Initialize an empty list $R$ for rank indices
\For{each token position $i$ in $T$}
    \State $input_i \gets$ Concatenate(previous tokens in $T$ up to position $i-1$)
    \State $token\_candidates\_pool_i \gets M(input_i)$ \Comment{Model predicts the next token candidates pool}
    \State $candidate\_set_i \gets$ Top $k$ tokens ranked by log probabilities from token candidates pool
    \State $actual\_token_i \gets$ Token at position $i$ in $T$
    \State $rank\_index_i \gets$ Find the rank of $actual\_token_i$ in $candidate\_set_i$
    \State Append $rank\_index_i$ to $R$
\EndFor
\State \textbf{return} $R$
\end{algorithmic}
\end{algorithm}

Initially, the first token of the text is input into the model, which then predicts the subsequent second token. The model outputs a candidate set containing all possible subsequent tokens along with their corresponding log probabilities. These candidates are also ranked in descending order based on their log probabilities. We then record the rank of the actual second token in the text within this candidate set. The process continues, the first two tokens of the text are input together to predict the third token. This step is repeated to generate a new candidate set and record the rank of the third token. This iterative expansion continues, with each iteration extending the length of the input text tokens to predict the next token and recording the rank of that token in its corresponding candidate set. 

\subsection{Watermark Detection}

In the watermark detection process, the primary goal is to ascertain the presence of a pre-defined signal watermark within the text generated by Large Language Models. The procedure involves analyzing the frequency spectrum of the text's token rank outputs and comparing it with the frequency characteristics of the embedded watermark pattern. The rank of each token in the generated text is computed with respect to the model's predicted candidate set, leading to a series of rank outputs, \( \text{rank\_outputs}_T \). Fast Fourier Transform is applied to the series of rank outputs, transforming the data into the frequency domain. The peak frequencies in the FFT spectrum are identified to uncover the embedded signal patterns. The peak frequencies, \( \text{peak\_freqs}_T \), are computed as:
    \begin{equation}
    \text{fft\_result}_T = \text{FFT}(\text{rank\_outputs}_T)
    \end{equation}
    \begin{equation}
    \text{peak\_freqs}_T = \text{FindPeakFrequencies}(\text{fft\_result}_T)
    \end{equation}
We use the sigmoid function to quantify this comparison, where the sigmoid function's input is the difference between the peak frequency of the watermarked text and the watermark pattern's peak frequency. If this value, after being processed through the sigmoid function, falls within a predefined threshold, it confirms the presence of the watermark. This comparison can be expressed as:

\begin{align}
    \Delta\text{freq} &= |\text{peak\_freqs}_T - \text{peak\_freqs}_{\text{pattern}}|
\end{align}

\begin{align}
    \text{is\_watermarked} = \sigma(-\Delta\text{freq})
\end{align}

\section{Experimental Evaluation}
\subsection{Research Questions}
\subsubsection{RQ 1. Watermark Detection Performance} We focused on embedding and detecting both single and multiple signal watermarks into text. First, to evaluate the effectiveness of embedding and detecting single signal watermarks in text. Furthermore, we extend our approach to detect and distinguish multiple signal watermarks to measure the capability of our method in differentiating among various signal watermarks.

\subsubsection{RQ 2: Impact on Quality of Model-generated Text} We investigate the trade-offs between watermark detection accuracy and the quality of model-generated text, as influenced by factors such as temperature, text length, and signal amplitude. Our investigation specifically focuses on how these parameters affect the perplexity of the generated text. During text generation, the temperature parameter adjusts the probability distribution of model-generated candidates, thereby influencing the top 5 candidate words list. Similarly, signal amplitude directly impacts token selection, affecting the PPL. Additionally, the length of model-generated text plays a crucial role in watermark detection. Texts that are too short may lack sufficient contextual information, leading to a decline in text quality and potentially compromising the effectiveness of watermark detection.

\subsubsection{RQ 3: Attacking Signal Watermark} We investigate the robustness of signal watermarks against various attacks and evaluate their effectiveness in distinguishing between human-written and watermarked texts. Our study focuses on four primary types of attacks: 1) The \textbf{Copy-Paste Attack} assesses the system's ability to distinguish between human-written and watermarked texts, with effectiveness measured by detection accuracy percentage. 2) In the \textbf{Substitution Attack}, we explore how substitution manipulations impact the watermark's strength. 3) The \textbf{Paraphrase Attack} involves examining the effects of paraphrasing on the watermark's robustness. 4) Additionally, the \textbf{Cross Watermarking and Detecting Model} aspect investigates whether the same model is required for both embedding the Signal Watermark and recomputing Token Probability to effectively detect the watermark's presence.

\subsection{Datasets}

In our experiment, mirroring the approach of Kirchenbauer et al. \cite{Kirchenbauer2023}, we utilized the 15G Realnewslike subset from the C4 dataset, as described by Raffel et al. \cite{Raffel2020C4}. This subset was chosen to simulate various realistic language modeling scenarios. We randomly selected texts and divided each into two segments: the first part served as the prompt, while the second part, formed by trimming a fixed number of tokens from the end of the text, was used as the baseline completion.

We gathered about 3000 samples from Realnewslike subset, ensuring each had a length at least 250 tokens. In scenarios of text generating, we suppressed the EOS ('\texttt{</s>}' or '\texttt{<|endoftext|>}') token to prevent the generation of overly short sequences. All generated sequences were then standardized to a length of 200 tokens. This method provided a uniform structure across different decoding techniques for effective comparison and analysis.

\subsection{Pre-trained Models and Benchmark}
In our research, we utilized open-source models from Meta, specifically OPT-1.3b and OPT-6.7b \cite{Zhang2022OPT}, along with OpenAI's text-davinci-003 \cite{td003}, as our foundational models. To measure the quality of the generated text, we employed perplexity as the metric, computed using the most advanced model, text-davinci-003. All our experiments were conducted using the OpenAI API and the Hugging Face library \cite{Wolf2019}. We introduce two model-generated text detectors that serve as benchmarks:

\textbf{DetectGPT} as introduced by Mitchell et al. \cite{Mitchell2023}, presents an innovative method for distinguishing between human and machine-generated text. This approach uniquely leverages the concept of probability curvature analysis, eliminating the need for specialized classifiers or the creation of datasets comprising either authentic or synthetic texts. Utilizing readily available mask-filling models such as T5 \cite{Raffel2020C4} and mT5 \cite{Xue2020}, DetectGPT evaluates the curvature within the probability distribution of a model's outputs.

\textbf{GPTZero} introduced by Tian et al. \cite{GPTZero}, stands out as a commercial AI detection tool, capable of discerning whether a piece of text has been produced by a large language model like ChatGPT. Its detection mechanism hinges on key metrics including perplexity and burstiness. Trained on an extensive dataset that includes both AI-generated and human-crafted English texts.

\subsection{Experimental Settings}
In our experiments, the default configuration, unless otherwise specified, involves utilizing the same model for both watermarking and detection tasks. The model's temperature parameter is set to 0 during text generation and Token Probability Re-computation. The signal amplitude, dictating the pattern range, varies from 1 to 5. We embed 10 tokens within a single-cycle signal pattern. When using the text-davinci-003 model, we configure it with a frequency penalty of 0, a presence penalty of 0, and a top-k setting of 1.

\subsection{Evaluation Metrics}
We utilize several metrics to assess the performance of our model in different scenarios:

    \textbf{AUC (Area Under the Curve):} This is used when evaluating with a balanced dataset comprising both human-written and model-generated text (watermark or non-watermark). AUC represents the ability of the model to distinguish between these two categories. A higher AUC indicates a better performance in differentiating human-written text from model-generated text.
    
    \textbf{FNR (False Negative Rate) and FPR (False Positive Rate):} These metrics are also employed for the balanced dataset. FNR measures the rate at which positive instances are mistakenly classified as negative (e.g., watermarked text wrongly identified as non-watermarked), while FPR indicates the frequency of false alarms (e.g., non-watermarked text incorrectly marked as watermarked).

    \textbf{ACC (Accuracy):} This metric is used when we are specifically evaluating the detection of watermarked texts. High accuracy in this context means the model is highly effective in correctly identifying watermarked texts.
    
    \textbf{Precision and Recall:} In the Attacking Signal Watermark section, we focus on detecting each token of the attacked text. To assess our model's performance in this scenario, we rely on two crucial metrics: Precision and Recall. During the Attacking phase, each token in the text is assigned labels, specifically 'Watermark' for watermarked tokens and 'Non-watermark' for other tokens. We implement a sliding window technique for watermark detection, enabling us to predict these labels for each token. \textbf{Precision} in this context quantifies how many of the tokens predicted as 'Watermark' are indeed watermarked. It measures the proportion of correctly identified 'Watermark' tokens out of all tokens identified as 'Watermark' (both correctly and incorrectly). A higher Precision indicates that the model is more precise in identifying 'Watermark' tokens, resulting in fewer false alarms (misidentified 'Non-watermark' tokens). \textbf{Recall}, on the other hand, measures the proportion of correctly identified 'Watermark' tokens out of all actual watermarked tokens. A higher Recall signifies that the model is effectively capturing most of the watermarked tokens, minimizing the number of missed ones.

\section{Experimental Results}
\subsection{RQ 1: Watermark Detection Performance}

\subsubsection{Single Signal Watermark Detection}

We construct a balanced dataset comprising a combination of human-written prompt from Realnewslike subset and watermark text generated by models. This setup allows us to evaluate the accuracy of our model-generated text detection method in comparison with benchmarks as DetectGPT and GPTZero. OPT and text-davinci-003 will produce two sets of watermark text, so we use their averages in the benchmarks calculation. The results are shown at Table \ref{tab:1}. We found that FNR of DetectGPT and GPTZero were high, which shows much of the text labeled as "human-written" by detectors was actually generated by the model. This is likely because our model-generated watermark text has a Perplexity (PPL) close to human-written text, and these tools rely on PPL differences for detection. These detectors typically rely on the PPL difference between human-written and model-generated text for detection. The table shows the effectiveness of our watermark detection approach.

\begin{table}[!ht]
\centering
\begin{tabular}{lccc}
\toprule
Model             & AUC    & FPR    & FNR    \\ 
\midrule
GPTZero           &0.64        &0.05        &0.86        \\
DetectGPT         &0.53        &0.04        &0.96        \\
\midrule
OPT-1.3b          &0.97        &0.04        &0.00        \\
text-davinci-003  &0.84        &0.24        &0.08        \\
\bottomrule
\end{tabular}
\caption{Watermark Detection Performance Comparison}
\label{tab:1}
\end{table}

Second, we design a balanced dataset consisting of No Watermark text and Watermark text. This is intended to gauge the sensitivity of the watermark detection. Table \ref{tab:2} shows the performance of OPT-1.3b and text-davinci-003 in detecting watermark versus non-watermark text. It suggests that while both models are effective in watermark detection in terms of accuracy and reliability.

\begin{table}[h]
\centering
\begin{tabular}{lccc}
\toprule
Model             & AUC    & FPR    & FNR    \\ 
\midrule
OPT-1.3b          &1.00        &0.00        &0.00        \\
text-davinci-003  &0.88        &0.24        &0.12        \\
\bottomrule
\end{tabular}
\caption{Sensitivity Analysis of Watermark Detection}
\label{tab:2}
\end{table}

\subsubsection{Multiple Signal Watermarks Detection}

We establish watermark patterns using different sinusoidal functions—specifically, sin(x), sin(2x) and sin(3x). We collected about 400 watermark text samples, embedding an average of three different signal watermarks. Each text segment is embedded with one of these three distinct watermark types. Our objective is to accurately differentiate three variations. Table \ref{tab:3} underscores the effectiveness of using sinusoidal functions for watermarking in texts generated by LLMs. The ability of both models to differentiate between sin(x), sin(2x), and sin(3x) with high accuracy is indicative of the robustness and precision of this watermarking approach.

\begin{table}[ht]
\centering
\begin{tabular}{lccc}
\toprule
Model & Acc (sin(x)) & Acc (sin(2x)) & Acc (sin(3x)) \\
\midrule
OPT-1.3b & 0.92 & 0.98 & 0.96 \\
text-davinci-003 & 0.87 & 0.89 & 0.92  \\
\bottomrule
\end{tabular}
\caption{Comparison of model accuracy for different watermarking signals.}
\label{tab:3}
\end{table}

\subsection{RQ 2: Impact on Quality of Model-generated Text}

\subsubsection{Impact of Models' Temperature}
\begin{figure}[!t]
    \centering
    \subfloat[OPT-1.3b]{ % Added description after \subfloat
        \includegraphics[width=0.8\columnwidth]{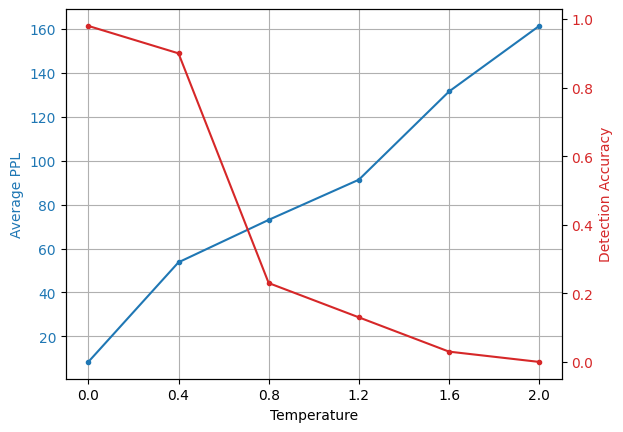}
    }
    
    \subfloat[text-davinci-003]{ % Added description after \subfloat
        \includegraphics[width=0.8\columnwidth]{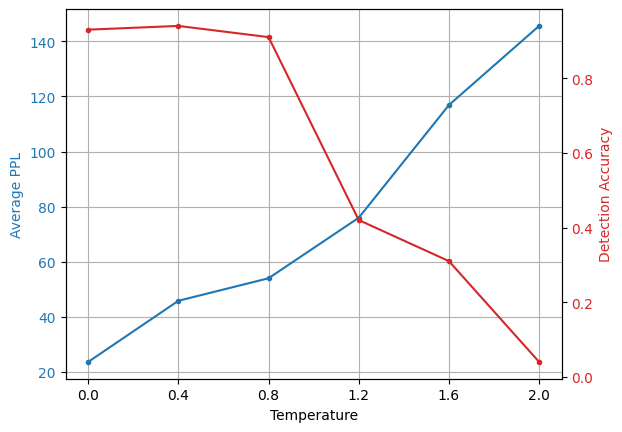}
    }
    \caption{Impact of Temperature on Watermark Detection Accuracy and Average PPL for two models: (a) OPT-1.3b (b) text-davinci-003. In scenarios with higher temperature settings, text-davinci-003 maintains a relatively high accuracy rate. In contrast, OPT-1.3b, while outperforming at lower temperatures, exhibits a more significant drop in accuracy as the temperature increases. This difference highlights the distinct response of each model to variations in temperature, particularly impacting their performance in watermark text detection.
}
    \label{fig:t}
\end{figure}

Although the model selects from these candidates based on a pre-defined pattern, the value of the temperature setting determines the relative probabilities of these words, thus influencing which words are more likely to appear in this token candidates pool. Figure \ref{fig:t} reveals how the detection accuracy of watermarked text and perplexity changes as the temperature parameter increases. As the temperature parameter increases, the detection accuracy of watermark text decreases for both models, and the text perplexity increases. But at the same time, a higher temperature can ensure the creativity of the text. The watermark text generated at a low temperature is similar every time. How to ensure the creativity of text and detection accuracy and PPL is a matter that needs to be traded off.

\subsubsection{Impact of Signal Amplitude}
\begin{figure}[!t]
    \centering
    \subfloat[OPT-1.3b]{ % Added description after \subfloat
        \includegraphics[width=0.8\columnwidth]{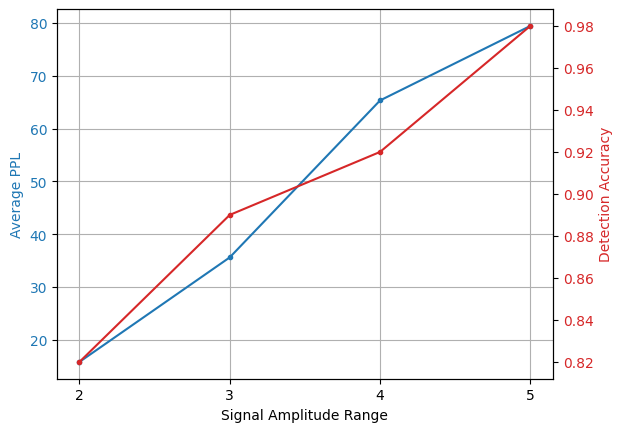}
    }
    
    \subfloat[text-davinci-003]{ % Added description after \subfloat
        \includegraphics[width=0.8\columnwidth]{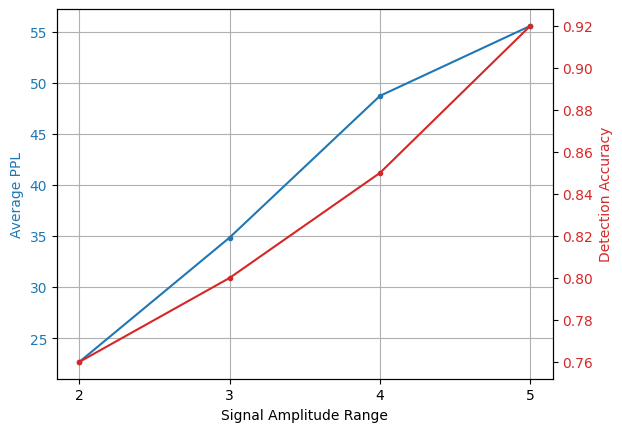}
    }
    \caption{Impact of Signal Amplitude on Watermark Detection Accuracy and Average PPL for two models: (a) OPT-1.3b (b) text-davinci-003. For both models, increasing the Signal Amplitude results in a rise in perplexity, thereby reducing text quality. However, this adjustment concurrently enhances the accuracy of watermark detection. This phenomenon illustrates a trade-off between text quality and the effectiveness of watermark detection influenced by Signal Amplitude.}
    \label{fig:amplitudes}
\end{figure}

Reducing the Signal Amplitude causes the model to select the next token from the top ranks of the token candidates pool, which decreases the perplexity and enhances the quality of the text. However, this simultaneously weakens the strength of the Signal Watermark, leading to a reduction in detection accuracy. Figure \ref{fig:amplitudes} shows the impact of increasing Signal Amplitude on watermark text detection accuracy and perplexity.

\subsubsection{Impact of Token Length}

\begin{figure}[!ht]
    \centering
    \includegraphics[width=0.8\linewidth]{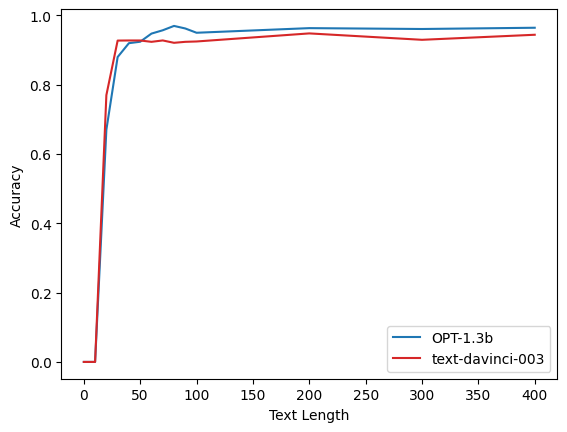}
    \caption{Detection Accuracy of Watermarked Texts of Varying Lengths. The prompt tokens that need to be used as context information differs by models, we can observe from Figure \ref{fig:len} that text-davinci-003 needs at least 20 tokens as prompt, and OPT-1.3b needs about 25 tokens as prompt.}
    \label{fig:len}
\end{figure}

We generate watermark texts of varying lengths using randomly selected prompts, creating 100 instances for each length. Subsequently, we detect these watermarks of different lengths and calculate the accuracy. Since Token Probability Re-computation is essentially a next token prediction process, a short token will cause the model to lack context information (i.e., prompt) when computing the log probability. As shown in Figure \ref{fig:len}, the performance will be affected in a short token scenario.

\subsection{RQ 3: Attacking Signal Watermark}
In the attack scenario, our detection approach incorporates a variation: we employ a 10-token sliding window that moves from left to right across the text, advancing one token at a time. Our focus is on re-computing token probabilities exclusively within this window during each pass. Any additional text to the left of the sliding window is used as contextual input to assist in recalculating these probabilities. As the window moves through the text, it generates a specific frequency value at each position. We compare this value with the frequency of the pre-defined watermark pattern. If the difference falls within a set threshold, it indicates the presence of a watermark in that section of the text. This method of analyzing the text token by token enables precise watermark detection at a granular level.

\textbf{Copy-Paste}: In this approach, we concatenate prompt texts from the Realnewslike subset with watermarked texts. The concatenated text is then analyzed to calculate the watermark detection Precision and Recall. Table \ref{tab:cp} presents these metrics for two models: OPT-1.3b and text-davinci-003. text-davinci-003 exhibits a higher precision in correctly identifying watermarked tokens with fewer false positives. Regarding recall, both models perform robustly, with OPT-1.3b achieving 0.98 and text-davinci-003 slightly surpassing at 0.99. This high recall reflects the models' efficiency in capturing almost all actual watermarked tokens, thus minimizing missed detections.

\begin{table}[ht]
\centering
\begin{tabular}{lccc}
\toprule
Model & Precision & Recall\\
\midrule
OPT-1.3b & 0.65 & 0.98 \\
text-davinci-003 & 0.84 & 0.99 \\
\bottomrule
\end{tabular}
\caption{Watermark Detection Performance of Models under Copy-Paste Attack.}
\label{tab:cp}
\end{table}

\begin{figure}[!t]
    \centering
    \subfloat[Substitution]{ % Added description after \subfloat
        \includegraphics[width=0.8\columnwidth]{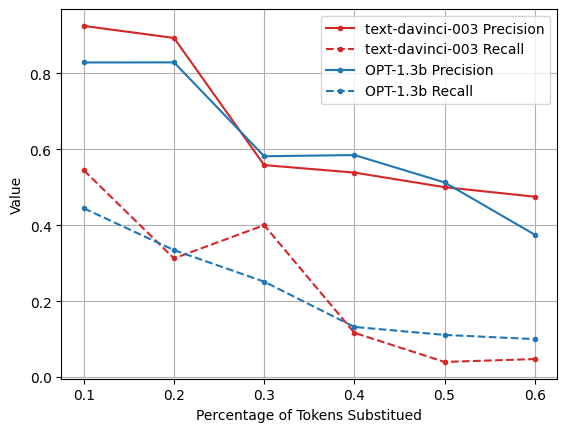}
        \label{fig:sub}
    }
    
    \subfloat[Paraphrase]{ % Added description after \subfloat
        \includegraphics[width=0.8\columnwidth]{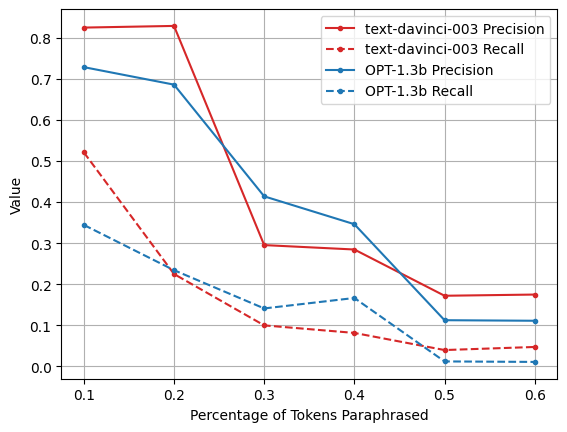}
        \label{fig:para}
    }
\caption{The Impact of (a) Substitution and (b) Paraphrasing on Watermark Detection Accuracy. This figure highlights that, despite a reduction in precision and recall with increasing Substitution and Paraphrasing levels, our watermark remains strong. Even when up to 60\% of the tokens are replaced or rephrased, we can still detect the presence of the watermark within the text segments. This durability is owed to the robust periodic signal features inherent in the watermark. It's important to note that Substitution and Paraphrasing can change the text's length, which might cause a discrepancy of about ±10 tokens in label alignment before and after these changes. Paraphrasing presents a greater challenge for watermark detection as it can change the text's semantic structure across continuous spans of tokens. This alteration has the potential to disrupt the embedded signal patterns more extensively than random substitution would.}
    \label{fig:subpara}
\end{figure}

\textbf{Substitution}: We use the 't5-large' model to progressively replace 10\% to 60\% of tokens in watermarked text, while concurrently attempting to detect the presence of watermarks with sliding window. The result is shown in Figure \ref{fig:sub}.

\textbf{Paraphrase}: We employ the 'T5ForConditionalGeneration' \cite{T5ForConditionalGeneration} method with the 't5-large' from the Hugging Face library for our task. This model is designed to handle English sentences, taking them as inputs and generating a collection of paraphrased versions. Paraphrased tokens span is set to 10. We also incrementally increase the proportion of paraphrased part within watermark text. By doing so, we aim to analyze and understand how such variations in paraphrasing affect the detection of watermarks in the text. The result is shown in Figure \ref{fig:para}.

\begin{figure}[!ht]
    \centering
    \includegraphics[width=\linewidth]{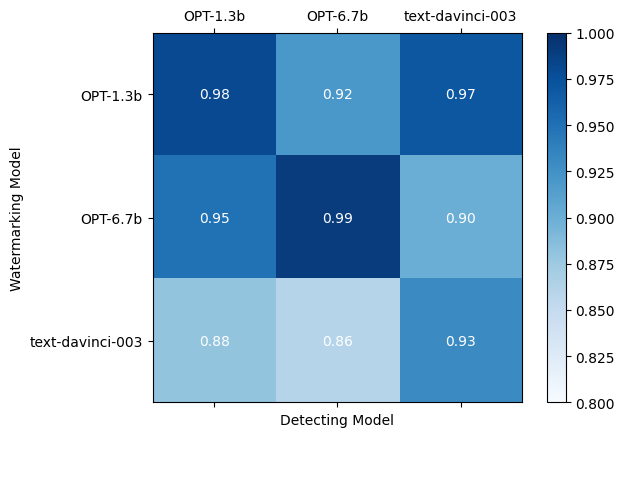}
    \caption{Evaluating Cross-Model Watermark Detection Accuracy in Large Language Models. White values are detection accuracy. The best performance is observed when the same model is used for both watermarking and detecting. This consistency suggests that each model is best at recognizing its own watermarking patterns, likely due to the specific nuances in text generation and encoding strategies inherent to each model. Both OPT-1.3b and OPT-6.7b demonstrate strong capabilities in detecting watermarks, regardless of which model was used for watermarking. When it comes to detecting watermarks embedded by other models, text-davinci-003 shows a marginally lower accuracy compared to the OPT models.}
    \label{fig:cross}
\end{figure}

\textbf{Cross Watermarking and Detecting Model}: We explore a cross-model approach using three different models: OPT-1.3b, OPT-6.7b, and text-davinci-003. This approach involves using each of the three models in both capacities — as a text generation model to embed watermarks (Watermarking) and as a Token Probability Re-computation model (Detecting). By interchanging these models, we aim to assess the effectiveness of watermark detection across different model architectures and training paradigms. Figure \ref{fig:cross} visualizes an experiment focused on cross-model watermark detection using three advanced language models: OPT-1.3b, OPT-6.7b, and text-davinci-003.

\section{Discussion}

Our system demonstrates the ability to detect watermarks within a remarkably short text span, requiring as few as 10 tokens, provided there's adequate context. This efficiency does not compromise the generated text's quality, which remains semantically and grammatically sound. Furthermore, our watermarking system proves effective across a temperature range from 0 to 1, despite temperature settings influencing token probability distributions.

In Section 3.1, we introduced an innovative signal watermark technique, showing the feasibility of using any periodic signal as an effective watermark. This flexibility enables the customization of signal watermarks to meet diverse needs, thus enhancing the method's versatility and practical utility.

Our experiments in Section 5.1.2 involved testing three different signal watermarks. Applying Nyquist's theorem \cite{Shannon1949}, we formulated a strategy for watermarking in textual content. For example, in a 100-token text, we assign every 10 tokens to represent a complete cycle of the watermark signal. This methodology aligns with the Nyquist criterion, enabling us to embed up to 50 unique signal watermarks (from sin(x) to sin(50x)) without compromising their integrity.

Building on this concept, we explored embedding sine waves into model-generated text, similar to modem signaling. This technique allows us to incorporate binary-coded information within these sine waves, effectively turning them into carriers of complex data. Utilizing periodic signals in this manner lets us embed meaningful information, like copyright details, directly into the text. This innovative approach merges advanced communication theory with text generation, heralding new possibilities for more secure and diverse digital watermarking applications.

\subsection{Limitations}
Despite the fact that the watermarked text is entirely generated by the model, it may still occasionally contain minor grammatical inaccuracies while remaining semantically invariant. Due to our strategy of circumventing the premature termination of text generation by disregarding stop symbols such as '\texttt{</s>}' or '\texttt{<|endoftext|>}', there may be instances of text redundancy. This implies that the model might generate text that is grammatically correct but contextually unrelated or superfluous. These issues could potentially impact the overall quality of the text. Enhancing the text quality and creative aspects remains an area for improvement.

Furthermore, in practical applications, the temperature parameter tends to be set slightly higher; for instance, the web version of ChatGPT typically operates with a temperature between 0.7 and 0.9. Addressing how to maintain the strength of the watermark and the quality of the text amidst variations in temperature represents a direction for our future research.

\subsection{Future Work}
In our future work, we aim to optimize the quality of text generated by our system. Our goal is to strike an ideal balance between the quality and creativity of the text and the robustness of the watermark. Additionally, considering the low information content in signals that indicate whether a text is model-generated, we plan to investigate the integration of error-correcting codes like Hamming codes \cite{Hamming} or Reed-Solomon codes \cite{Reed}. These codes, combined with carrier modulation techniques such as Quadrature Phase Shift Keying \cite{Proakis}, could be used to transmit textual information in ASCII code. This strategy would allow the model-generated text to carry a greater amount of information.

Furthermore, we are looking to expand our watermarking techniques to a wider array of AI-generated content (AIGC), including code snippets, AI-generated images, and videos. This expansion would diversify the application of our signal watermarking method, making it applicable to various forms of digital content created by AI.

\section{Conclusion}
In this study, we introduced an innovative watermarking method - Signal Watermarking - tailored for use with Large Language Models (LLMs). Our approach is grounded in a critical hypothesis about LLMs and employs signal sampling patterns as a benchmark for embedding signals into text. Throughout the process of embedding and detecting watermarks in LLM-generated text, we integrated and applied concepts from signal processing. The inherent properties of the signals ensured the strength and sensitivity of the watermarks, enabling them to remain effective across various attack scenarios and be detectable within very short spans of text. Our method demonstrated high accuracy in watermark detection, surpassing most existing open-source and commercial detectors in identifying model-generated text. This approach is versatile and adaptable across different types of LLMs, enabling the generation of watermarked text even with access limited to model APIs. The success of our technique illustrates the potential of combining signal processing with text generation, paving the way for new applications in safeguarding digital content authenticity and integrity in the AI era.

\vfill

\newpage
\appendix

\section*{A. Comparative Outputs from Models with and without Watermarks at Different Temperature Settings}

\begin{figure*}[!ht]
    \centering
    \includegraphics[width=\linewidth]{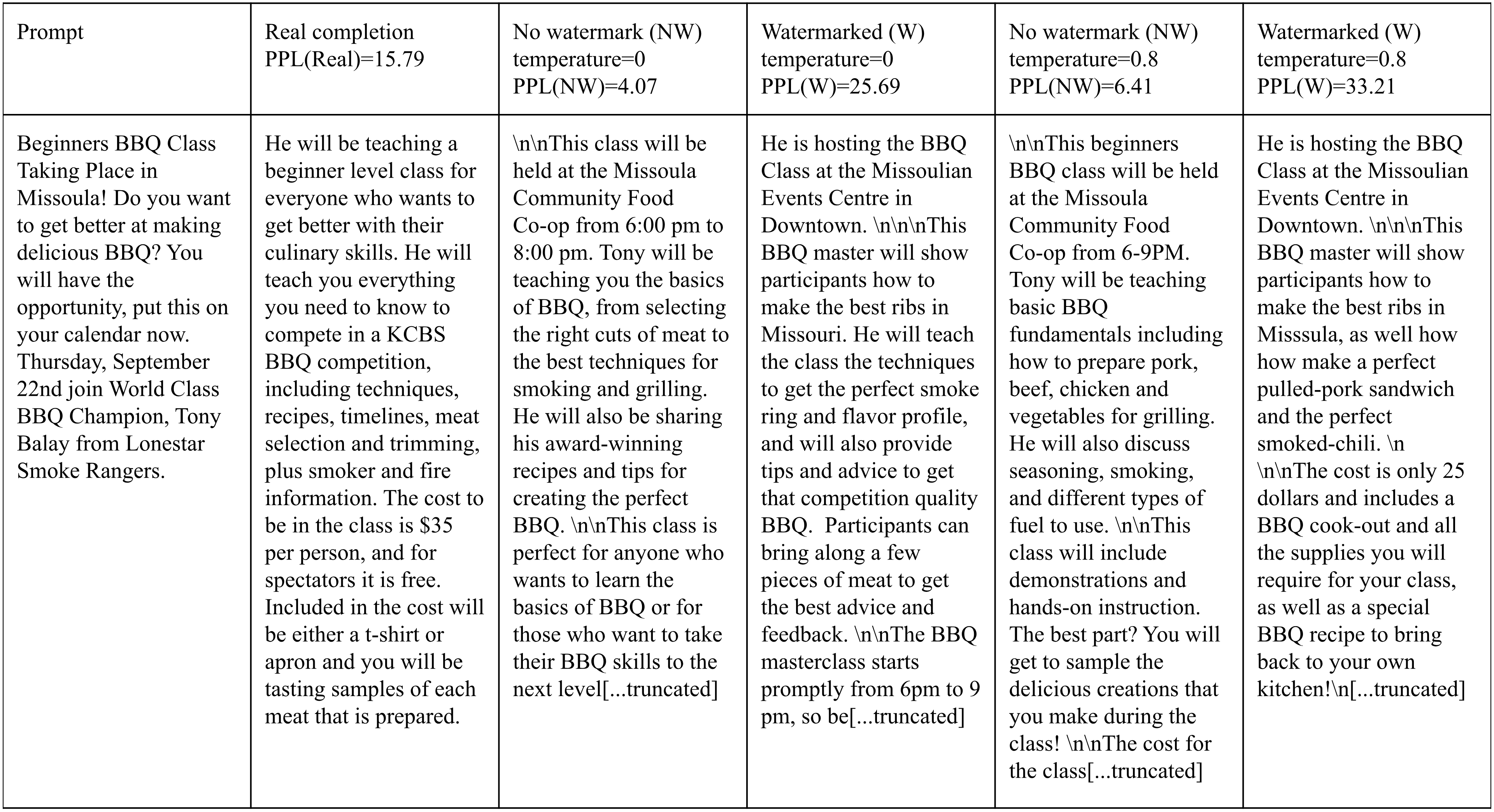}
    \caption{Comparison of Outputs at Temperatures 0 and 0.8 Using Prompts and Real Completions from the C4 Dataset. Both non-watermarked and watermarked texts were generated by the text-davinci-003 model. The Perplexity (PPL) values were also calculated using text-davinci-003.}
    \label{fig:app}
\end{figure*}

This appendix presents a detailed analysis of the Perplexity (PPL) scores obtained from texts generated by the text-davinci-003 model under varying temperature settings and watermarking conditions. The data includes comparisons between non-watermarked (NW) and watermarked (W) texts, as well as human-written text (Real completion), derived from the C4 dataset. The example comprises four different text, each generated at specific temperature settings (0 and 0.8) with and without watermarking. The PPL scores for these texts are calculated using the text-davinci-003 model. The data is summarized as Figure \ref{fig:app}.

\subsection*{B. Analysis and Observations}

The data indicates a clear influence of temperature settings on the PPL of generated texts. For both NW and W texts, an increase in temperature from 0 to 0.8 results in a rise in PPL, suggesting heightened unpredictability or complexity in text generation under higher randomness. The NW text at a lower temperature (0) exhibits a significantly lower PPL compared to the human-written text, implying a higher level of predictability or coherence. Conversely, the watermarked text at the same temperature shows a higher PPL than the human-written text, indicating the additional complexity introduced by the watermark. At the higher temperature (0.8), both NW and W texts have higher PPL scores than the human-written text, with the watermarked text demonstrating a substantially higher PPL. The consistently higher PPL scores for watermarked texts, as compared to their non-watermarked counterparts, highlight the impact of watermark embedding on the text's complexity. This trend is observable across both temperature settings.


\begin{thebibliography}{1}
\bibliographystyle{IEEEtran}


\bibitem{chatgpt} chatgpt. Available at: \url{https://chat.openai.com/}

\bibitem{radford2018improving}
Radford, A., Narasimhan, K., Salimans, T., \& Sutskever, I. (2018).
Improving language understanding by generative pre-training.
Retrieved from \url{https://s3-us-west-2.amazonaws.com/openai-assets/research-covers/language-unsupervised/language_understanding_paper.pdf}

\bibitem{radford2019language}
Radford, A., Wu, J., Child, R., Luan, D., Amodei, D., \& Sutskever, I. (2019).
Language models are unsupervised multitask learners.
OpenAI Blog.
Retrieved from \url{https://cdn.openai.com/better-language-models/language_models_are_unsupervised_multitask_learners.pdf}

\bibitem{brown2020language}
Brown, T. B., Mann, B., Ryder, N., Subbiah, M., Kaplan, J. D., Dhariwal, P., ... \& Amodei, D. (2020).
Language models are few-shot learners.
\textit{arXiv preprint arXiv:2005.14165}.
Retrieved from \url{https://arxiv.org/abs/2005.14165}
\bibitem{FFT}Cooley, J. W., \& Tukey, J. W. (1965). An algorithm for the machine calculation of complex Fourier series. Mathematics of Computation, 19(90), 297-301.

\bibitem{Xue2020}Xue, L., Constant, N., Roberts, A., Kale, M., Al-Rfou, R., Siddhant, A., ... \& Raffel, C. (2020). mT5: A massively multilingual pre-trained text-to-text transformer. arXiv preprint arXiv:2010.11934.
\bibitem{Wolf2019} Wolf, T., Debut, L., Sanh, V., Chaumond, J., Delangue, C., Moi, A., ... \& Rush, A. M. (2019). Huggingface's transformers: State-of-the-art natural language processing. arXiv preprint arXiv:1910.03771.
\bibitem{Raffel2020C4} Raffel, C., Shazeer, N., Roberts, A., Lee, K., Narang, S., Matena, M., ... \& Liu, P. J. (2020). Exploring the limits of transfer learning with a unified text-to-text transformer. The Journal of Machine Learning Research, 21(1), 5485-5551.

\bibitem{Devlin2019BERT} Devlin, J., Chang, M. W., Lee, K., \& Toutanova, K. (2019). BERT: Pre-training of deep bidirectional transformers for language understanding. arXiv preprint arXiv:1810.04805.
\bibitem{Liu2019RoBERTa} Liu, Y., Ott, M., Goyal, N., Du, J., Joshi, M., Chen, D., ... \& Stoyanov, V. (2019). RoBERTa: A robustly optimized BERT pretraining approach. arXiv preprint arXiv:1907.11692.
\bibitem{Radford2019GPT2} Radford, A., Wu, J., Child, R., Luan, D., Amodei, D., \& Sutskever, I. (2019). Language models are unsupervised multitask learners. OpenAI Blog.
\bibitem{Brown2020GPT3} Brown, T. B., Mann, B., Ryder, N., Subbiah, M., Kaplan, J., Dhariwal, P., ... \& Amodei, D. (2020). Language models are few-shot learners. arXiv preprint arXiv:2005.14165.
\bibitem{Zhang2022OPT} Zhang, S., Roller, S., Goyal, N., Artetxe, M., Chen, M., Dewey, C., ... \& Kiela, D. (2022). OPT: Open pre-trained transformer language models. arXiv preprint arXiv:2205.01068.
\bibitem{Smith2022LLaMA} Smith, N. A., \& Team, Meta AI. (2022). Introducing LLaMA: The Large Language Model Meta AI. Meta AI Blog.

\bibitem{T5ForConditionalGeneration} T5ForConditionalGeneration. Available at: \url{https://huggingface.co/transformers/v3.0.2/model_doc/t5.html}

\bibitem{Shannon1949} Shannon, C. E. (1949). Communication in the Presence of Noise. Proceedings of the IRE, 37(1), 10–21. https://doi.org/10.1109/jrproc.1949.232969

\bibitem{Hamming}Hamming, R. W. (1950). Error detecting and error correcting codes. The Bell system technical journal, 29(2), 147-160.
\bibitem{Reed}Reed, I. S., \& Solomon, G. (1960). Polynomial codes over certain finite fields. Journal of the Society for Industrial and Applied Mathematics, 8(2), 300-304.
\bibitem{Proakis}Proakis, J. G., \& Salehi, M. (2008). Digital Communications (5th ed.). New York, NY: McGraw-Hill.
\bibitem{Abdelnabi2021} Abdelnabi, S., \& Fritz, M. (2021, May). Adversarial watermarking transformer: Towards tracing text provenance with data hiding. In 2021 IEEE Symposium on Security and Privacy (SP) (pp. 121-140). IEEE.

\bibitem{He2022} He, X., Xu, Q., Zeng, Y., Lyu, L., Wu, F., Li, J., \& Jia, R. (2022). CATER: Intellectual property protection on text generation APIs via conditional watermarks. Advances in Neural Information Processing Systems, 35, 5431-5445.

\bibitem{Zhao2022} Zhao, X., Li, L., \& Wang, Y. X. (2022). Distillation-resistant watermarking for model protection in NLP. arXiv preprint arXiv:2210.03312.

\bibitem{Kirchenbauer2023} Kirchenbauer, J., Geiping, J., Wen, Y., Katz, J., Miers, I., \& Goldstein, T. (2023). A watermark for large language models. arXiv preprint arXiv:2301.10226.

\bibitem{Zhao2023} Zhao, X., Wang, Y. X., \& Li, L. (2023). Protecting language generation models via invisible watermarking. arXiv preprint arXiv:2302.03162.

\bibitem{Peng2023} Peng, W., Yi, J., Wu, F., Wu, S., Zhu, B., Lyu, L., ... \& Xie, X. (2023). Are You Copying My Model? Protecting the Copyright of Large Language Models for EaaS via Backdoor Watermark. arXiv preprint arXiv:2305.10036.

\bibitem{Yoo2023} Yoo, K., Ahn, W., Jang, J., \& Kwak, N. (2023, July). Robust multi-bit natural language watermarking through invariant features. In Proceedings of the 61st Annual Meeting of the Association for Computational Linguistics (Volume 1: Long Papers) (pp. 2092-2115).

\bibitem{Li2023} Li, L., Jiang, B., Wang, P., Ren, K., Yan, H., \& Qiu, X. (2023). Watermarking LLMs with Weight Quantization. arXiv preprint arXiv:2310.11237.

\bibitem{Gehrmann2019} Gehrmann, S., Strobelt, H., \& Rush, A. M. (2019). GLTR: Statistical detection and visualization of generated text. arXiv preprint arXiv:1906.04043.

\bibitem{Solaiman2019} Solaiman, I., Brundage, M., Clark, J., Askell, A., Herbert-Voss, A., Wu, J., Radford, A., Krueger, G., Kim, J. W., Kreps, S., et al. (2019). Release strategies and the social impacts of language models. arXiv preprint arXiv:1908.09203.

\bibitem{Mitchell2023} Mitchell, E., Lee, Y., Khazatsky, A., Manning, C. D., \& Finn, C. (2023). DetectGPT: Zero-shot machine-generated text detection using probability curvature. arXiv preprint arXiv:2301.11305.

\bibitem{Guo2023} Guo, B., Zhang, X., Wang, Z., Jiang, M., Nie, J., Ding, Y., ... \& Wu, Y. (2023). How close is ChatGPT to human experts? Comparison corpus, evaluation, and detection. arXiv preprint arXiv:2301.07597.

\bibitem{GPTZero} GPTZero. Available at: \url{https://gptzero.me/}
\bibitem{Writer} Writer: AI Content Detector. Available at: \url{https://writer.com/ai-content-detector/}
\bibitem{AITextClassifier} AITextClassifier. Available at: \url{https://freeaitextclassifier.com/}
\bibitem{Copyleaks} Copyleaks: Plagiarism Detector. Available at: \url{https://copyleaks.com/}
\bibitem{Sapling} Sapling AI. Available at: \url{https://sapling.ai/}
\bibitem{td003} text-davinci-003. Available at: \url{https://platform.openai.com/docs/models}

\bibitem{K}Kirchenbauer, J., Geiping, J., Wen, Y., Katz, J., Miers, I., \& Goldstein, T. (2023, July). A watermark for large language models. In International Conference on Machine Learning (pp. 17061-17084). PMLR.

\end{thebibliography}
\end{document}